# e-Commerce Business Models in the Context of Web 3.0 Paradigm

Fernando Almeida[1], José D. Santos[2] and José A. Monteiro[3]

[1]Faculty of Engineering, University of Porto, Porto, Portugal
[2]Polytechnic Institute of Porto, School of Accounting and Administration, Porto, Portugal
[3]National Institute of Systems and Engineering of Porto, INESC TEC, Porto, Portugal

## ABSTRACT

*Web 3.0 promises to have a significant effect in users and businesses. It will change how people work and play, how companies use information to market and sell their products, as well as operate their businesses. The basic shift occurring in Web 3.0 is from information-centric to knowledge-centric patterns of computing. Web 3.0 will enable people and machines to connect, evolve, share and use knowledge on an unprecedented scale and in new ways that make our experience of the Internet better. Additionally, semantic technologies have the potential to drive significant improvements in capabilities and life cycle economics through cost reductions, improved efficiencies, enhanced effectiveness, and new functionalities that were not possible or economically feasible before. In this paper we look to the semantic web and Web 3.0 technologies as enablers for the creation of value and appearance of new business models. For that, we analyze the role and impact of Web 3.0 in business and we identify nine potential business models, based in direct and undirected revenue sources, which have emerged with the appearance of semantic web technologies.*

## 1. INTRODUCTION

The new tendencies on the information concepts created by the Internet transformed a simple communication act into an important component of business and everyday life. Internet quickly change from a limited set of services, like navigation on simple web pages, or reading email, to a very broad and complex services based on connectivity that contextualizes people in space and time. A great contribution become from the miniaturization of the hardware communication devices and its increasing computational capabilities. To be more precise, we cannot be sure if the growth of the Web has influenced by the miniaturization of the hardware devices, or if the hardware device miniaturization came accelerate the growth of the Web. In fact, seem to have mutual influence. In parallel with the miniaturization of the hardware devices, the emergence of sub-communities on the Web, known as social-networks, has accelerated people communication needs. The communications across the Web are no longer limited to the email or a one to one perspective. There are a wide variety of options to an individual to exteriorize ideas, expressions, emotions, and thoughts, in real time and in a selective way. Each individual are free to choose which information share, when and with who wants to share. Nevertheless, email does not decrease importance in the face of such range of resources. By the contrary, in today's, email is more like an id that identifies uniquely an individual on the Web. Also, email as become the hub the other services use to link an individual to an electronic profile.

Due different stages of evolution, Web was categorized according different versions. The stage that corresponds to Web 3.0 are seeing as the stage where web provides the necessary conditions for individuals and organizations to use information in ways that facilitate the exchange of content, independently of the devices and the networks. Evidences show us that in recent times have increase the use of mobile devices, namely smartphones. This factor cannot be dissociated from facts that show people using their cell phone to read news while take a coffee, or share a picture take from the cell phone during a visit to a museum, or publish its state on Facebook while in a rock concert, or even rank a food restaurant while take a meal. This new phenomena that lead people to use their mobile devices all the time, since they wake-up until to get sleep, should begin anticipating by organizations and prepare to capture value.

Business models are necessarily influenced by users' behaviour in the use of the new technologies. After ups and downs of the dotcom companies, electronic business has rediscovered new opportunities through the massive use of mobile devices and the increasing of mobile apps. At the same time, with the increase importance of semantic web, interoperability is becoming a philosophy and a relevant issue in a business strategy. First movers companies look to know more about their customers to get more information about its preferences and interests. Social networks become community hives that can be easily broadcast a product, a new concept or even get new ideas from users' feedback. However, some work need to be done. More than collect tags, cutting individual sentences or quotes, it is necessary to work on it. Web 3.0 provides contextual and semantic conditions and the human creativity and technique do the rest.

In this paper we contextualize the main characteristics of Web 3.0 and frame with e-Commerce concept. In the final chapters are identified the sources of value in semantic Web, followed by business models strategies.

## 2. FUNDAMENTAL CHARACTERISTICS OF WEB 3.0

The technological changes of the concepts on the World Wide Web have a tendency to be classified according version numbers. After a period over 10 years of World Wide Web implementation and adoption, some regulations have been introduced. Additionally the evolution of Web technologies have contributed to emerged the first version, Web 1.0. To the second version, Web 2.0, it was not necessary to wait a long period. In the present, people already talk about Web 3.0 [21].

It was not known the real facts to Internet have been versioned, perhaps because of the close relation with computer science and programming technologies. About the past versions it seems to have some acceptance about its main characteristics. However, scholars and experts do not seem to agree about versioning neither on a specific set of characteristics or technologies to define clearly each version, namely version 3.0. In fact, definitions about Web 3.0 are far from consensus. Experts point semantics and higher personalization as most relevant features. Ubiquity is also an important characteristic [21]. Common people can interact anytime and anywhere, independently of the framework, operating system or application, using a common virtual space known as Internet. Evidences points that the major contribution for this high human connectivity through computational devices become from: (i) the performance of the computational devices; (ii) the advances on telecommunications technologies, namely wired communications; (iii) the miniaturization of powerful devices, that allow communicate with high mobility; (iv) the increasing development of "intelligent" programs, named "agents" to automate a large and varied number of operations in the Web; (v) and the broad acceptance of this new paradigm by the human users.

…International Journal of Advanced Information Technology (IJAIT) Vol. 3, No. 6, December 2013

The "visible" part to common sense, are social networks. These spaces can be seen as segmented networks in the World Wide Web network. Inside of each social network, more segmentation can be identified. For example: friends groups, professional groups, colleagues groups, family groups, learning groups, business groups among many others that already exist in real world context. The ways of these are formed and how they self-adjust and what information produces is a phenomenon that as already motivated the academic community attention.

A big question still remains: Are these aspects enough to clearly characterize Web 3.0?
In our approach we consider that Web 3.0 characterization frontiers are not completely defined. Nevertheless, regarding the objective of this paper, Web 3.0 can be identified by a set of characteristics that are changing the way of how people interact beyond the physical world and have influence of how make business: (i) Ubiquity - people can be connected anytime and anywhere. A multiplicity of communication options like mobile networks, Wi-Fi networks, cable networks or fibber networks in addition with different types of devices like laptops, desktops, tablets and an infinity set of mobile devices are facilitator to stay connected. No matter what system or architectures uses. As refer Wolfram (2011) Web 3.0 is where the computer is generating new information [12]; (ii) Individualized - information can be segmented and contextualized by individual interests and by each network contact. In fact, a feature of Web 3.0 is the capability to deal with unstructured information on the Web more intelligently providing contextual meaning to the published information regarding individual or group variables [1]. A known form of connection with a very wide set of adopters are the social networks. These "large hubs" allow people to decide, about his own visibility and which are their contacts based on information interests and "proximity". Proximity, in this context, should be understood as synonym of friendship, school colleague, work colleague, client, friend of a friend, family, among other concepts of social relationships; (iii) Efficient - Information can be filtered by context, significance and relevance by humans and by computational devices. This characteristic is implicitly related with the individualized characteristic. Regarding the differences among individuals, information produces different reactions in different people. The capability of filter information according individual interest it seems to be a value-add of Web 3.0. Efficiency implies that the information has to be structured in such a way that machines can read and understand it as much as humans can, without ambiguity [6].

Perceiving the potential the networks and the concept of proximity to spread information, organizations look to Web 3.0 as an alternative to get more close of the potential clients. Content websites and collaborative websites become more iterative and facilitators to new business models. A first example is how information is publicized. Usually, each contact can assume to main roles: (i) to be a "relay" to pass the word; and (ii) to be a feedback provider. However, people maintain the possibility to filter what is important to him-self. For example: when a user of Facebook subscribes information about a certain brand of mobile phones, the information will be feed by this user. The network friends (contacts) of the user will know that he (she) has subscribed new information about mobile phones. However, they can decide to subscribe or they can filter if they do not consider relevant. On the other hand, the organizations can explore the network and the preferences of its "fans" (subscribers) to send personalized information to attract their interests.

Another example are advertises. Websites are no longer merely hub contents. More than that, it provides advertising oriented to user interests and change dynamically. For example, when a user visits a news website he (she) may face publicity spots in context with the category of the news. Another example is "free" webmail services like Google mail. When a user opens each email message, adverts are in context with the email content.

33

A cross perspective of Web 3.0 is provided by Fuchs et al. [7]: Supported on social theories of Karl Marx about community building and collaborative forms of cooperation, characterizes Web 3.0 as a system of human cooperation [7]. In fact Web tends to be socio-technical system where human interactions are mediated by technologies and those technologies assume an increase role of represent the system more close of reality [7].

## 3. THE ROLE OF WEB 3.0 IN BUSINESS

Decision-making is one of the great challenges of management. In their daily lives managers make decisions whose amplitude ranges with associated different consequences. Its action unfolds in an operational, tactical or strategic plan. To reduce the risk of the decision, the quantity and quality of available information plays a decisive role.

Along the years Web has taken on greater prominence as a source of information for managers. It is possible to obtain information about the products marketed by the clients (Web 1.0) and also know what the belief that customers have of products (Web 2.0). We passed from read-only to read-write paradigm. All these developments made grow the amount of information available and increased the difficulty of managers in filter the important facts, and in structuring information coming from different sources and in various formats.

The Web 3.0 is the answer to the contextualization of existing information by specific user need [1, 16]. The information provided will also depend on the resources held by the service provider and on the characteristics of devices used to access the Internet [16].

The Web 1.0 connects pages with information, the Web 2.0 connects people and the Web 3.0 is about linking and connecting web of data but transforming it in knowledge. With the web 3.0 is expected that search engines produce different results by the user: one user, one question, one result according to her profile. The web 3.0 will organise and assembles the pages found in a search engine, by themes, topics. The idea is to read, analyse and identify the directions of semantically words so as to relate the information to each other. Additionally it will also be able to deal with the interests of people previously defined by the user, or uses his/her profile previously built in an application that will allow its use by other. In addition, the browsers themselves over time analyse the main interests of the person and use it to improve the quality of searches. The more we research, the more the browser learns (i.e. a web more intelligent and allowing increasingly user profile is powered). It will act as if it were an assistant who register their interests as the user navigates and "serves" information, making suggestions according to what we know, the historic. This use of the user profile allows companies that communicate via the web address a more targeted message most likely to return.

In a business perspective, the construction of the Web 3.0 is based on three pillars: the datastore and provision of services in the cloud; social media and user-generated content to add personal or perspectives that provide value to third parties; migration to a protocol that will facilitate the device connectivity (IPv6) [22].

The move towards Web 3.0 will bring opportunities for the growth of companies operating in the area of information technologies, as organizations in general will feel new needs updating or redesigning their web presence. Increasingly the customization of content presented will be based on a solid user profile refine that can be stored on the computer of his own or resorting to other entities that provide this service. The perfecting of customization will allow a better user experience with web page that interacts.

With the Web 2.0 it is expected that this create rich, meaningful user experiences, allowing people to easily locate and collaborate with others, support the quick and easy creation of content, store and leverage information to create value. The web 3.0 will add features that will allow the consumer to benefit from vastly personalised experience; context-aware, precise response; efficient management of time spent on the Web; much more personalized Web experience [19]. On the other hand, there will also be some benefits for the company: great opportunity for rethinking strategies and business processes, better targeted marketing; improvement in operational efficiency and reduction of costs [22].

With the Web 3.0 trust and privacy reach a greater role, enabling the identity management in order to ensure the authenticity and the user education promoting good practices of Internet use, may be growing business areas [22].

Business flows should be reformulated in such a way as to promote connectivity and automation, promoting the relationship with the customer and providing mechanisms which adapt to their reality and contemplate different types and sources of data [22]. Some work has been done on this area, namely by fast food chains and similar companies. This can be experienced in cities such as London or Paris, where stores offering free wi-fi to their customers can be found. For free access to the service, customers fill out a small registration providing personal information like their name, email or even birth date. This task is performed merely once. In the next visit to the same or to another store of the same chain, customers will be identified as soon as they connect to the corresponding wi-fi hotspot, by providing their credentials. The immediate information benefits for business are: to know how many different customers visit their stores; how much time a customer spends in each store; how many stores have been visited by the same costumer; which is the frequency of visits per store; the possibility to direct campaigns and marketing; the possibility to direct questionnaires to evaluate customer satisfaction or product preferences.

The main goal of the stores is not to provide access to the Internet or process the information generated by its customers while using the Internet. As so, they only control a part of the process: the offer of the wi-fi service. The other part of the process is in the hands of the emerging companies which main purpose is to offer wi-fi access.

In summary, these companies create the conditions to control the access and traffic between stores wi-fi hotspots and Internet. Additionally they can process statistical information and delivery to the main companies that own the store chains. An example is the Cloud that operates in a few countries of Northern Europe, namely United Kingdom.

Web 3.0 is not only an opportunity for improve business, but also has become an opportunity to create new business.

## 4. E-COMMERCE FOR WEB 3.0

The Web is currently part of modern society, allowing store, access and share information anytime, anywhere. The human being has been presenting increased concerns with their education, leisure, mobility and health, so there is an opportunity here for web 3.0 develop, charting in products such as vehicles, support equipment to health, clothes and even in points of sale and the packaging of products [21]. Furthermore, the effect of geo-referencing already allows a geographical customization framed along with user profile [22].

The web 3.0 will provide the link between objects, which will facilitate the emergence of new services and the mobility of people. In the other hand, it tends to organize and assembly the pages by themes, topics and interests previously expressed by the user, which enables a more effective advertising. [22]

Advertising on the Web is also a form of business. The presentation of ads, search result for certain words can come to develop. The words assume an increasingly important role in the message that is displayed, with a triangle composed of word search, word on the user profile, word associated with the communicational announcement.

The retail companies can adapt better their offerings to clients, optimising cross-selling and improving the customer shopping experience, taking also advantage of integration with social networks and geo-referencing. The Web 3.0 integration is between power machines and platforms, having as work centre the content and its significance [19, 22]. However, to the customer care the output, i.e., how the company manages to convey customer value [20], this value should be added by the Web's potential. For the creation of value the company may seek to monetize the location of the user, device used, day of the week, time of day and weather.

One of the aspects that cannot be neglected is the possibility of making purchases without leaving the social networking. For example, the Facebook already offers several Apps for integrate many storefront applications into Facebook business Page. Now, we can go shopping without leaving the social networking (Social Shopping). With Web 3.0 it is expected that the content of these storefront are provided taking into account all its characteristics, including geo-referencing, profile, history, customization of the product, specific payment conditions. The Web 3.0 encourages one-to-one instead of the one-to-many not only on communicational perspective, but also in the marketing of products [8].

Media companies and entertainment area can adapt more easily and customize in more detail their products and services using their customers' online experience [19, 22]. But the mass customization using an integration with semantics are a concern to meet the individual needs increasingly looking not to lose any economies of scale and bearing in mind the need to make smaller and smaller series, ideally unitary.

The brands with reputation tend to have a strong presence on the web for the engagement of various users, seeking to monetize their presence especially in social networks. Thus, it is natural that we feel need to harness the potential of Web 3.0 in integrating information from this, all the knowledge accumulated through other communication channels or from other points of sale. The Web 3.0 not unfolds from the point of view of business, just around the Internet, but also has new challenges in terms of multichannel integration which provides a 360-degree view of the customer. The challenge, possibly, will not only bring to Web information from various points of interaction. We must also define how traditional outlets can incorporate technology that provide a more customized shopping experience and monetize all the semantic web 3.0 explored.

The CRM 3.0 (Customer Relationship Management) leverages the potential SCRM (Social Customer Relationship Management) adding the advantages of semantics Web. Social CRM builds upon CRM by leveraging a social element that enables a business to connect customer conversations and relationships from social networking sites in to the CRM process. The focus is now on community and relationship building via social venues (e.g., Facebook and Twitter). Communication is no longer based only in business-to-customer, but also customer-to-customer and customer-to-prospect have an important role. Customers collaborate with businesses directly or indirectly to improve products, services, and the customer experience. Furthermore, conversation is less formal and more "real" moving from brand speak to community speak. As a consequence, there is thus an opportunity for the emergence of new services, new applications that integrate this broader version of CRM.

The sites that offer comparison of products and prices are another challenge for companies. With Web 3.0 the comparison of competition may not be carried out only under the same or similar products, but in all the products that are framed in the same context, consistent with the profile of the user and that may go against future needs taking into consideration a possible evolution of the consumer.

Regardless of the product marketed, the remuneration model of e-commerce, or the parties involved, the Web 3.0 provides challenges for the suppliers, which must always bear in mind the concept of client with unique needs, the proposed products according to the accumulated know-how about each client leveraging new forms of interaction and of technologies that provide for continuous learning.

## 5. BUSINESS MODELS STRATEGIES

The concept of business models has been brought to the forefront of strategic thinking. It has become an important factor through advances in information and communication technologies, in particular on the Internet and broadband technologies [2, 9]. According to Hawkins [10] a business model can also be seen as the "architecture of a business". In fact, a business model is about how a certain strategy or goal can be realized. The term business model is commonly used in both science and practice. However, there are multiple definitions on what a business models actually consists of.

Two of the most popular and integrated approaches were introduced by Chung et al. [5] and Osterwalder & Pigneur [17]. For Chung et al. [5] the business model is a new approach to evaluate an organization that makes possible the identification of several elements, such as resources, processes, products, clients and activities; Osterwalder & Pugneur [17] defines a business model as a conceptual tool that contains elements and their relationships, allowing to define the business logic of a company. In these approaches there are several commons central elements in the business model, respectively: who is the consumer?, what gives value to the consumer?, how to create and capture value?, what is the economic reason that explains how to deliver value to the consumer at an appropriate price?.

At the same time, with the increase importance of semantic web, interoperability is becoming a philosophy and a relevant issue in a business strategy. Interoperability cannot be reached just collecting or cutting individual sentences or quotes, but its concept includes having access to customer data before other people do and seeing relationships in that data which connection and relations aren't immediately apparent. Companies look to know more about their customers to have a better position in selling a product that interests them. If a company knows what the customers like to do, where they like to go, and who they like to spend time with, placing targeted advertisements and define a value proposition to them becomes easier.

Some business models appear from the potentialities of semantic web technologies and Web 3.0. These business models are based in a range of direct and indirect revenues. Additionally, a large amount of these business models propose a central role for the customers and open developer community, which turn them active elements in the process of value acquisition. In the following, we will detail the main business models that offer a good representation of the different ways in which organizations can monetize, directly or indirectly, from Web 3.0 platforms.

### 5.1. Licensing

The licensing model is already very stable in the software industry [18]. The idea in the Web 3.0 environment is charge fees to let developers use data in other environments. In this way, the application and the use of the data inside the application domain is totally free of charge and only the data used outside the application has a cost for the developer.

This is a well known model that with Web 3.0 will be improved as the data is easier for people to have new applications with newly combined data [4]. The buyers wish that their suppliers change their products or services to incorporate what they really need in particular. One of the ways to build demand is to modify our products and sell more to our current clients. One other way is offer and sell products that are modified for new types of clients [13].

### 5.2. Subscription

The subscription can be seen as variation of the licensing model where a subscription period is included. The subscription model charges for access to data for a period of time. Rather than sell the service directly, the promoter of the service is selling weekly, monthly or yearly access to the service or data contents. This, in effect, converts a one-time sale of a product into a recurring sale of a service.

The implementation of a subscription model requires the definition of the specific terms of use for the offered product or service and the limits of their usage. For instance, some subscription business models base their rates upon how often or for how long a customer uses the service. A good strategy is to offer long-time memberships with lower prices, which will create sufficient capital up front that the company can use for financial investments and further research and development. Additionally, it would be interest to guarantee that the renewal process is effortless. This may involve automatic renewals, where the customer will be charged once the original term of usage expires unless the user specifically designates otherwise. Anyway, this can have potential legal issues in many countries and the company must be aware that many customers may find this unethical. Therefore, it is important to make it clear during the initial contract.

### 5.3. Premium Services

The idea behind this business model is to offer simple and basic services for free for the user to try and more advanced or additional features at a premium. For these companies, growth comes not only from new customers, but more often from existing ones. If clients are happy with the free portion of a service, they are substantially more likely to pay a small fee to get additional services. More importantly, they are the ones who evangelize a start-up business on social networks, as well as tout it to friends and family. Furthermore, the customers who trust the product so much they are willing to pay are even more inclined to refer to their friends and family.

The premium services are a complement to the free service that helps the company to be known to the consumer. This has to be developed to take full advantage of semantic web. The behaviour of each user has to be the jumpstart for the companies offer. This can help the consumers' choice because it's offered only what the user looks for.

This model allows the users to use a part of the product and interact in a way to create demand for other services or simple more added services provided at a established cost. This information can be obtained from the users in forms and forums at no cost.

### 5.4. Advertising

At the present time, most social media services contain advertisements. Ads by Google are the most widespread and popular business model [23]. The idea in the Web 3.0 world is to charge a small number of advertisers for brand visibility of sponsoring the data. Getting charges for ads placed around data on web pages can be an important source of revenue.

Essentially there are two approaches to implement an advertising promotion model in a Web 3.0 platform. Firstly is to see it as a web portal. Therefore, a high volume of user traffic makes advertising profitable and permits further diversification of site services. Secondly we can use some advanced business intelligence techniques as contextual advertising and behavioural marketing. Therefore, the ads are not static and can dynamically change due to the data content or the behaviour of user in the Web 3.0 platform. As a consequence, contextual advertisers can sell targeted advertising based on an individual user's surfing activity.

Advertisement can gain a new life with this platform and consumers will have more choice in the offer they will receive. By receiving only the information that is relevant to the consumer it is more likely that he will pay more attention than when he is being overflowed with information that doesn't suits the consumer in question.

Depending in how significant the traffic is, advertising can have lower or higher costs or profits. The implementation of this model is made easier with the Web 3.0 platform by directing the advertisers to the users that really search for the product. This has lower costs and offers more efficiency.

### 5.5. Traffic Generation

The traffic generation business model is a good example of an indirect monetize source of revenue. The idea is very simple and consists in publishing data to earn favourable positions in search engines and other directories to generate more traffic. When the company gets more traffic, then there is a big chance to increase the revenues from other direct business models like e-payments, premium services or advertising.

Viral traffic is something that everybody would like to achieve but it's not as simple as it looks. The use of social network like Facebook or Twitter, the use of YouTube has to have a well established strategy. This strategy doesn't necessarily have costs. The ideas that have bloomed with this model have been simple. Having accomplished high traffic generation, a site can then use other model like advertising and the affiliate program more effectively. The feedback on this model action is great in recognition, business offers and sales.

### 5.6. e-Payments

The idea is to get revenues from micro-payments for the individual use of data sets. These micro-payments can be easily deployed using electronic payments, namely using credit cards, e-checks, online bill pay and paypal.

Electronic payment is very convenient for consumers. Consumers can make e-payments at any time of the day or night, from just about anywhere in the world. At the same, it is a very fastest, simplest and safest way to perform a bill transaction. On the other side, electronic payments lower costs for businesses. The more payments they can process electronically, the less they spend on paper and postage. Offering electronic payment can also help business improve customer retention. A customer is more likely to return to the same Web 3.0 site where his or her information has already been entered and stored.

The e-Payments model is not new in the Internet. However, the evolution to subscription or micro-payment model will be a relevant change to Internet content. It will change the Internet experience and could significantly impact lower income users. There will still be a free content that will allow people to have a limited number of free views per month in an attempt to interest new subscribers.

### 5.7. Mass Customization

Mass customization lets business deliver individualised products to every single customer, based on the buyer's choice of aesthetic, functional, or contextual components such as styles, colours, materials and measurements. The individualised tailoring of products has, traditionally, been too costly to scale. However, mass customization allows customers to participate meaningfully in the design of their goods, restoring individually to the process, while leveraging the cost-efficiencies of mass production to make the model feasible.

The mass customization allows customers to get the exact products they want. Simultaneously, by offering goods that cannot be found elsewhere, the model enables companies to differentiate themselves and build more engaged and loyal relationships with customers. Additionally, by getting commitment and payment at the top of purchase funnel, mass customizer decreases the transitional business risk.

Blecker et al. [3] considers that the main downside of mass customization is the increased cost and complexity of production, which is ultimately reflected in higher prices. A good strategy to avoid this issue is the identification of key elements of a product to make customizable and offering the right degree of variability in order to provide a made-to-order feeling, while ensuring manageable and scalable production.

### 5.8. Collaborative Consumption

Collaborative consumption is considered the next generation of sharing, trading and renting in a different activity sectors. It has gain a bigger importance with the growth of peer-to-peer marketplaces and other technology-enabled platforms. Using this model, sellers can get rid of used or depreciating assets, while buyers can consume the items' residual value at a price that is substantially cheaper than retail. Furthermore, collaborative consumption can also includes rental models, which are suitable for high cost items.

The collaborative consumption is based in the principle of peer-to-peer (P2P) models. Plenty of companies on the Web are turning to P2P, but so far, few are making a profit. According to Knapp [11] despite big user numbers generated by P2P firms, it is difficult for them to profit from what is essentially considered free file sharing. In fact P2P models face obvious challenges around trust and quality control. First, presence of P2P networks raises an important question for firms in terms how piracy cannibalizes products sales or actually can enhance sales. Second, P2P networks also provide companies and individuals with a new low cost tool for legitimate distribution of content that can be used for competitive advantage.

P2P networks could become a significant component of the digital strategy for content providers in a variety of sectors. However, for many of these business models to succeed, there is a need to understand and properly evaluate how copyright holders use digital rights management (DRM) techniques to complement P2P technologies, how can online communities complement P2P distribution models, and how can security and authentication be assured in a distributed distribution environment [14].

### 5.9. Mobile Commerce

Mobile-commerce takes traditional e-commerce models and leverages emerging new wireless technologies to permit mobile access to the Web. Wireless technology can be used to enable the extension of existing Web business models to service the mobile work force and consumer of the future [15]. Wireless networks utilize newly available bandwidth and communication protocols to connect mobile users to the Internet. The key technologies used are cell phone-based 3G and 4G, Wi-Fi, and Bluetooth.

The Mobile Web 3.0 has elements that build upon prior eras of Web 1.0 and Web 2.0, but it also has several distinct and different elements from what's come before. Some of these distinct elements of Mobile Web 3.0 era include real-time time, ubiquitous platform, location aware and sensors.

## 6. CONCLUSIONS

The evolution to Web 3.0 brings new challenges and potential for business in a wide range of fields. The Web 3.0 concept integrates relevant search, location-based services, mobile enablement and rich social interaction in a single online experience. Web 3.0 dramatically enhances the user experience and offers new ways of interaction between vendors, consumers and suppliers. As a consequence, companies need to adjust their business strategy in order to take advantage of the big wave of data that is available by the seamless integration of different platforms and tools.

Semantic web technologies and Web 3.0 brought new business models to the market. Some of them are based in direct revenues where the access to the service is directly being paid by the users, such as licensing, subscription, e-payments or mobile commerce. On the other side, there are more indirect revenue sources, where the service is indirectly being paid by traffic or market reputation, such as advertising, traffic generation or collaborative consumption.

As future work it would be interested to identify and measure the most relevant business models in the context of Web 3.0. For that, we expect to make an international statistical study in order to collect different opinions from chief executive officers about the potential of each identified business model. Besides that, we also want to measure the level of maturity of each business model.

## Authors

**Fernando Almeida** is a professor and IT consultant at ISPGaya and Faculty of Engineering of University of Porto. He holds a PhD. in Computer Science Engineering from Faculty of Engineering of University of Porto (FEUP). He has more than 10 years of experience working for IT companies and research centres. His research interests include innovation policies, entrepreneurship, supporting decision systems and big data management.

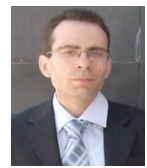

**José D. Santos** is a professor at ISPGaya and School of Accounting and Administration from Polytechnic Institute of Porto. He is also a marketing consultant for CRM and SCM projects. He holds a MSc. in marketing and he is attending the PhD in Management. His research interests include marketing, customer relationship and advertising.

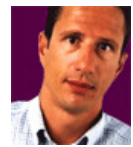

**José A. Monteiro** is a professor at ISPGaya and researcher at INESC TEC. He holds a MSc. on Information Management from Faculty of Engineering of University of Porto (FEUP) and he is attending the PhD. in Informatics Engineering (ProDEI) at FEUP. His research interests include information management, social web and semantic technologies.

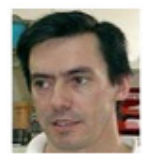